\documentclass{webofc}
\usepackage[varg]{txfonts}   
\usepackage[section]{placeins} 
\usepackage{natbib,hyperref}

\begin{document}

\title{Third-party transfers in WLCG using HTTP}

\author{
        \firstname{Brian} \lastname{Bockelman}\inst{6}\fnsep\thanks{\email{bbockelman@morgridge.org}} \and
        \firstname{Andrea} \lastname{Ceccanti}\inst{3} \and
        \firstname{Fabrizio} \lastname{Furano}\inst{1} \and
        \firstname{Paul} \lastname{Millar}\inst{2} \and
        \firstname{Dmitry} \lastname{Litvintsev}\inst{5} \and
        \firstname{Alessandra} \lastname{Forti}\inst{4} 
}

\institute{European Organization for Nuclear Research (CERN) \and 
Deutsches Elektronen-Synchrotron (DESY) \and 
Istituto Nazionale di Fisica Nucleare (INFN) \and
University of Manchester \and
Fermi National Accelerator Laboratory \and
Morgridge Institute for Research
}

\abstract{%
Since its earliest days, the Worldwide LHC Computational Grid (WLCG) has relied on GridFTP to transfer data between sites. The announcement that Globus is dropping support of its open source Globus Toolkit (GT), which forms the basis for several FTP client and servers, has created an opportunity to reevaluate the use of FTP. HTTP-TPC, an extension to HTTP compatible with WebDAV, has arisen as a strong contender for an alternative approach.

In this paper, we describe the HTTP-TPC protocol itself, along with the current status of its support in different implementations, and the interoperability testing done within the WLCG DOMA working group’s TPC activity. This protocol also provides the first real use-case for token-based authorisation for this community. We will demonstrate the benefits of such authorisation by showing how it allows HTTP-TPC to support new technologies (such as OAuth, OpenID Connect, Macaroons and SciTokens) without changing the protocol. We will also discuss the next steps for HTTP-TPC and the plans to use the protocol for WLCG transfers.
}

\maketitle
\section{Introduction}

\begin{figure}
    \centering
    \includegraphics[width=10cm,keepaspectratio]{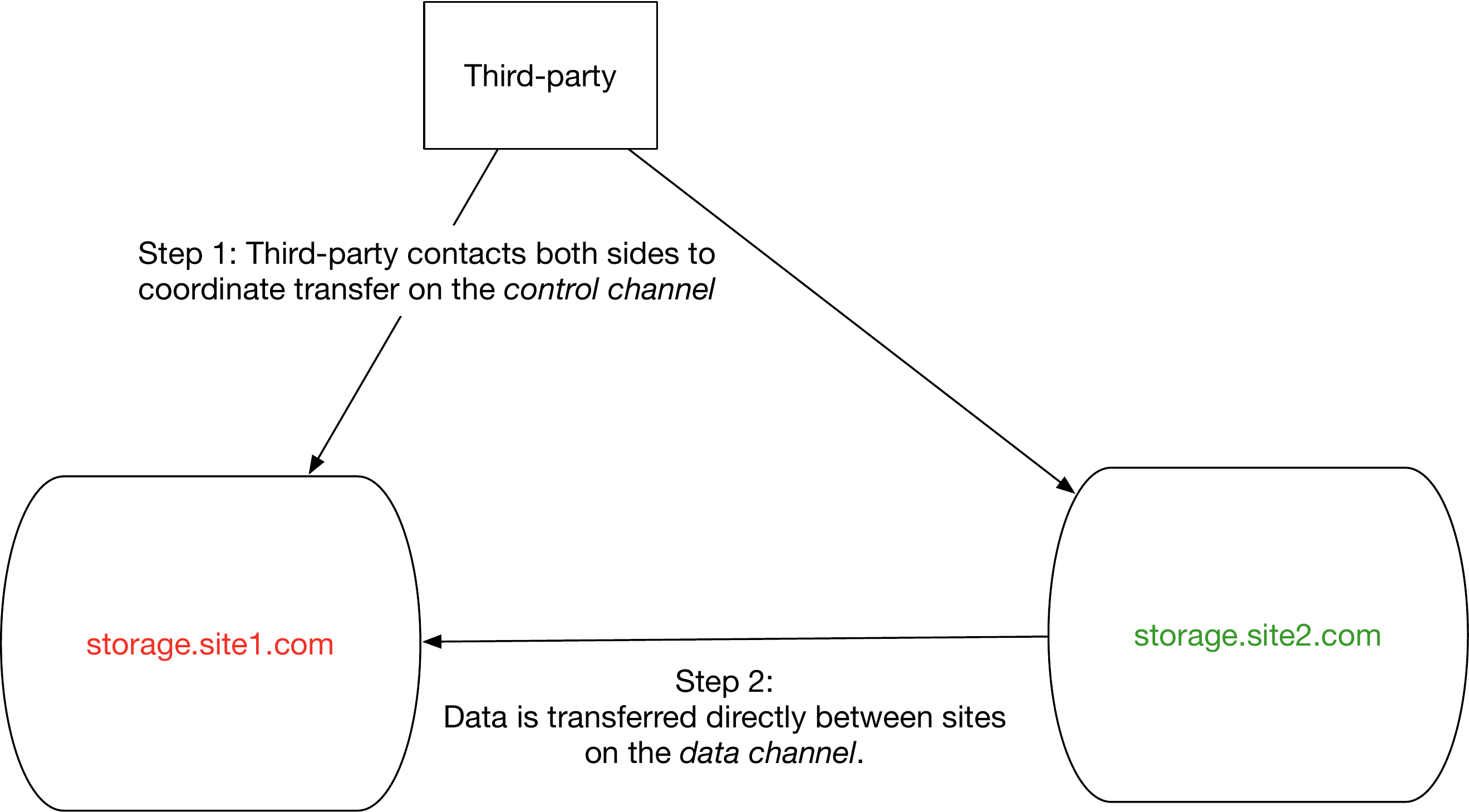}
    \caption{The basic concept behind TPC: an entity independent of either storage service coordinates direct transfer of data between the source and destination.}
    \label{fig:tpc}
\end{figure}

The primary driver for wide-area data movement for all LHC experiments is bulk data movement between storage services.  This bulk data movement serves to pre-stage data to be processed by production systems or to increase data replication to make it more available for analysis.  The technique to perform these transfers is third-party copy (TPC); in TPC, a central entity (the `third party') contacts a source and destination storage endpoint to facilitate a transfer from the source to the destination.  This provides for central management and coordination of transfers but allows for data to move directly between the storage systems.  The high-level concept is illustrated in Figure \ref{fig:tpc}.

In 2017, Globus announced the retirement of the Globus Toolkit, which served as the reference implementation for GridFTP protocol \cite{gfd.20, gfd.21}; this has increased interest into a number of alternatives such as HTTP \cite{rfc2068}.  HTTP is protocol that underpins the World Wide Web, making it one of the most common protocols on the planet - meaning there is a large community of experts and many mature implementations.  Unlike GridFTP, utilizing HTTP does not expose the WLCG community to the risks of relying on a specialized protocol.

The initial work to adopt HTTP as a third party transfer protocol within the WLCG community was outlined in \cite{bootstrapping}, following activities descripted in \cite{httpecosystem}; since then, the protocol has evolved and matured into what we term ``HTTP Third Party Copy'' (\texttt{HTTP-TPC}), as described in Section \ref{sec:background}.  Further, the WLCG community has formed a working group around Third Party Copy as part of the \textit{Data Organization, Management, and Access} (DOMA) initiatives; as outlined in Section \ref{sec:community}, this working group is testing and developing both HTTP-TPC and third party copy extensions for XRootD \cite{xrootd-tpc-prot}.  This has allowed for the continued growth of the activity - both in maturing implementations so they can be used in production and further evolving the protocol as described in Section \ref{sec:evolutions}.

\section{Background \label{sec:background}}

In preparation for the Run~1 of the LHC, a number of transfer protocols were considered by the LHC community before it eventually settled on GridFTP with a reference implementation provided by the Globus Toolkit \cite{globus_gridfp} (at least one other production-quality implementation has been written by the dCache project \cite{dcache_gridftp}).  This transfer protocol was augmented with the Storage Resource Management (SRM) protocol \cite{srmv2.2} which helped manage load-balancing between servers and the storage end-points.  Overall, the GridFTP protocol has served the community faithfully for nearly 15 years.

By the end of Run 2, several events transpired that motivated the community to re-evaluate its use of GridFTP as a TPC protocol.  First, many sites began to retire their SRM endpoints as unique space management features of SRM were largely never used, GridFTP could be used directly and native load-balancing solutions were introduced.  Second, the Globus organization's retirement of the Globus Toolkit \cite{globus-support} meant the implementation of GridFTP  in use by several of the storage systems had no original developer support.  This led to the formation of the WLCG DOMA TPC working group during the WLCG DOMA face-to-face at CHEP 2018, charged with examining alternative options and growing nascent ecosystems.

For the HTTP-TPC, as explained in \cite{bootstrapping}, the key concept is the use of the WebDAV \texttt{COPY} verb.  The client sends an HTTP request using \texttt{COPY} to the \textit{active endpoint} of the transfer along with another URL in an HTTP header.  For \textit{pull mode}, the active endpoint is given a \texttt{Source} header; for \textit{push mode}, the active endpoint is given a \texttt{Destination} header.  The active endpoint then downloads or uploads, respectively, from the \textit{passive endpoint}. Features of note for the HTTP-TPC protocol include:

\begin{itemize}
\item Separation between the “framing” and the “transfer” protocol.  The URL sent to the active endpoint does not have to use \texttt{https://}; for instance, the dCache implementation has shown that the HTTP-TPC active endpoint can be given a GridFTP URL to move the data over GridFTP (potentially useful for enabling a transition from GridFTP).
\item The active endpoint sends continuous performance markers back to the TPC client, allowing the client to monitor progress (cancelling the transfer as necessary).
\item Ability for transfers to be load-balanced using HTTP’s built-in redirection response.
\item When in ‘pull’ mode (the active endpoint is the destination), multiple pipelined \texttt{GET} requests can be load-balanced across multiple parallel TCP streams, allowing a single transfer to proceed faster compared to when a single TCP stream is used.
\end{itemize}

As of March 2020, there are four independent implementations of the HTTP-TPC protocol, in the dCache, DPM, StoRM~\cite{storm_webdav}, and XRootD software products.  Further, as XRootD often forms the basis of other storage services in the WLCG community, services like EOS also have HTTP-TPC support without needing a separate implementation. 

\begin{figure}
    \centering
    \includegraphics[width=10cm,keepaspectratio]{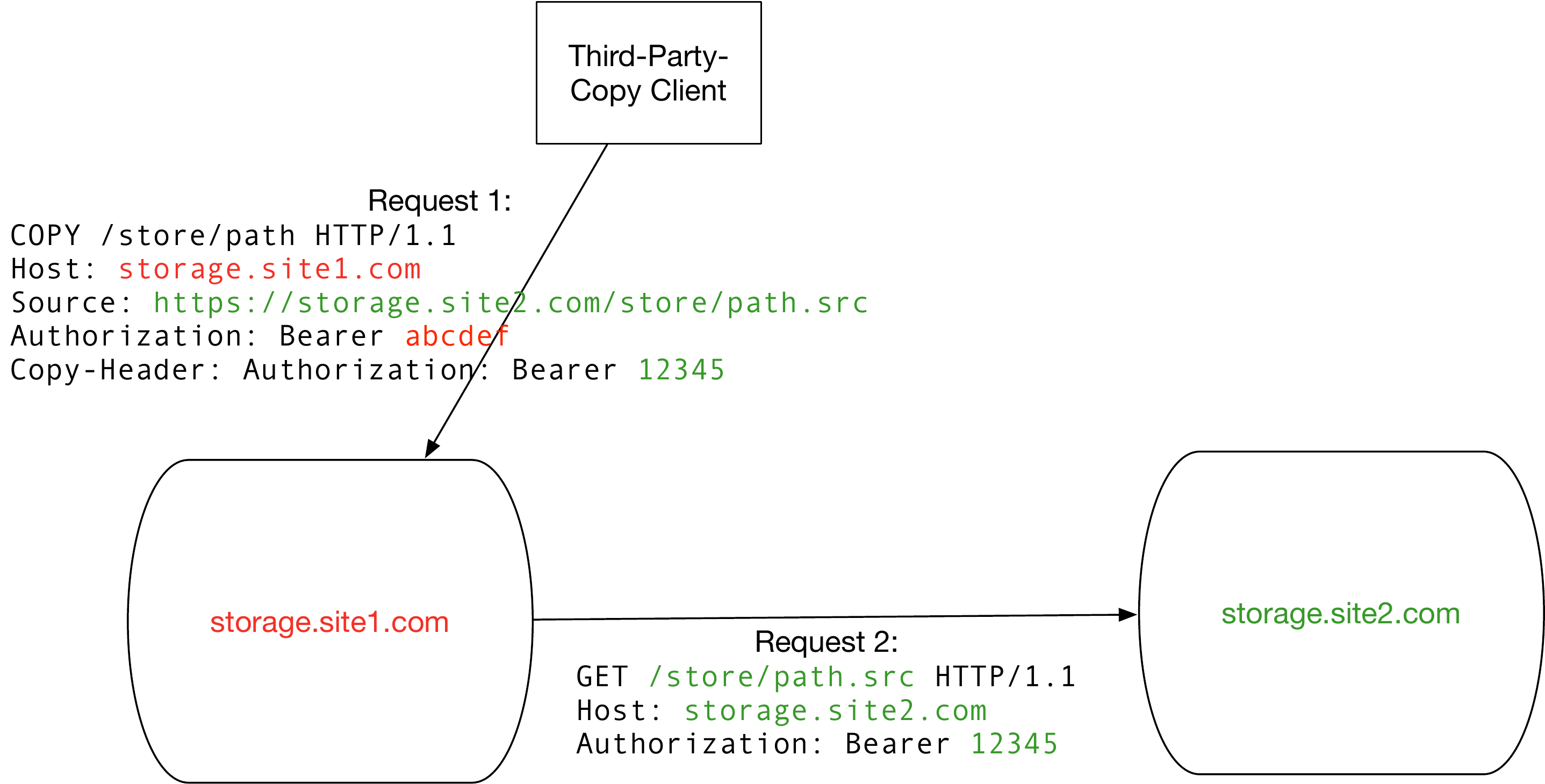}
    \caption{The basic mechanism of HTTP-TPC, reproduced from \cite{bootstrapping}.  Here, we illustrate ``pull mode", where the third party copy client contacts the destination site and issues a request that the destination downloads (\textit{pulls}) the data from the source over HTTP.}
    \label{fig:http-tpc}
\end{figure}

\section{Building the HTTP-TPC Community \label{sec:community}}

We found that the key to maturing the use of HTTP-TPC beyond initial specification and implementation is to build a user community.   The primary mechanism thus far has been the WLCG DOMA TPC working group; in the 18 months that followed the work done in \cite{bootstrapping}, this group (with co-leads Bockelman and Forti) has coordinated the development and finalization of HTTP-TPC, helped deploy a test-bed for HTTP-TPC, and organized a testing infrastructure. Within the testing infrastructure, the working group operates three types of tests: nightly, integration, and full scale.

The nightly “smoke tests” \cite{smoke-tests}, demonstrate compliance and functionality with the HTTP-TPC protocol by performing a small transfer against a known working endpoint in addition to simple tests for acquiring a transfer token from the endpoint (Section \ref{sec:evolutions}).  These tests are driven by a simple script and are meant to be easily reproducible by developers or administrators.

As of March 2020, the HTTP-TPC `smoke test' test-bed has 49 participating endpoints.  To catch bugs and issues as early in the development process as possible, we include endpoints from across the full software development life cycle: from endpoints on production sites to integration test-beds to developer instances.  This approach allows us to test across multiple versions (e.g., stable releases transferring against the latest nightly builds) in addition to across multiple implementation.

The continuous integration tests the ‘transfer matrix’ between any two protocol endpoints in the system at a small-scale (2-3 GB per hour), allowing the working group to understand behavior at a modest scale and to monitor for a broader set of pairwise issues.  These are driven by a dedicated instance of Rucio \cite{rucio}, the data management solution used by the ATLAS experiment.  Unlike the smoke tests, these integration tests include the full stack of WLCG transfer utilities. While failures are more difficult to reproduce in this environment, these tests are far more representative of a production transfer activity.  For example, failures due to the interaction of the FTS3-based \cite{fts3} TPC client and the storage service may require a full FTS3 server to reproduce.

\begin{figure}
    \centering
    \includegraphics[width=10cm,keepaspectratio]{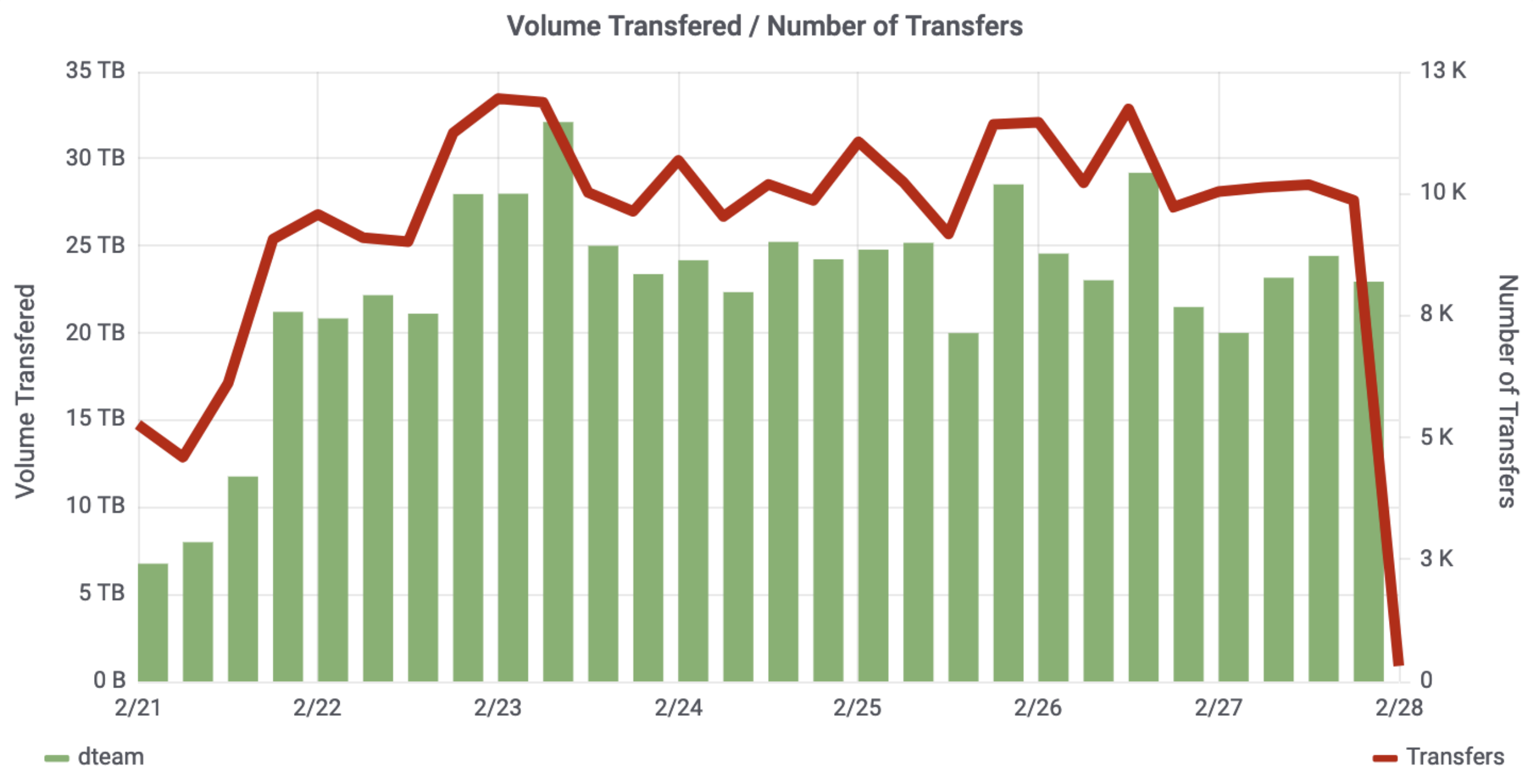}
    \caption{Example graph from the CERN monitoring system displaying the number and volume of HTTP-TPC transfers in the scale test-bed per 6 hour interval.  Note that maintaining 20TB of data movement every 6 hours for a week adds up to 560TB of data transferred.}
    \label{fig:scale}
\end{figure}

Finally, full-scale tests are also driven by Rucio; other than scale, these are identical to the continuous integration tests.  These scale tests are driven by a driver script that uploads a randomly-generated one terabyte dataset to an endpoint, then generates a rule to replicate the dataset to all other endpoints in the system.  Once the dataset is completely transferred, the driver script triggers a deletion of the replicated datasets; after the deletion is complete, the same rule is installed again.  With this setup, we have demonstrated the ability to transfer approximately half a petabyte of data per week.  See Figure \ref{fig:scale}.  This allows us to monitor for issues at scales similar to the production system.

\section{Evolving Approach \label{sec:evolutions}}

One of the strongest motivations for using HTTP as the base protocol in TPC is that it allows for a number of authorization schemes.  The WLCG has historically used the Grid Security Infrastructure (GSI) \cite{gsi} with various extensions; at its core, GSI is based on X.509 PKI.  As GSI is based on X.509, its infrastructure can be used to associate a TLS session for HTTPS and authenticate a client and the TLS level.  Previously in \cite{gridsite}, work was done to delegate a grid proxy to the HTTPS-based active endpoint; with the client's delegated identity, the active endpoint could then authenticate its transfer request to the passive endpoint.

However, this use of GSI suffers from the same issue as GridFTP: the retired Globus Toolkit was the reference implementation.  The WLCG DOMA TPC working group has instead settled on using bearer tokens to authenticate transfers.  A number of bearer token based schemes have been proposed, including SciTokens \cite{scitokens} and the WLCG Common JWT profile \cite{wlcg-jwt}; these will eventually allow a transfer to be performed completely without the use of a X.509 client credential.  As a transition mechanism, we have defined a way for clients to request a token from the storage endpoint, provided the request is made over a HTTPS connection that is GSI-authenticated.

First, the client must perform OAuth2 metadata discovery \cite{rfc8414} against the storage endpoint to determine the associated (storage-specific) token endpoint.  Then, an access token request is made against the token endpoint using the client credentials flow \cite{rfc6749}.  Unlike a typical client credential flow, when the client authenticates via an HTTP header (such as the \texttt{Authorization} header), this request must be done over the GSI-authenticated HTTPS channel.  Based on this token request and the client's authorization, the storage endpoint will issue an access token that can be used as part of the HTTP-TPC infrastructure.  Although the token format is considered opaque (implementations have been done both based on JWT \cite{rfc7519} and Macaroons \cite{macaroons}), the client must ask for one or more of an agreed-upon set of scopes of the form \texttt{\$ACTIVITY:\$PATH}.  If permitted, the returned token will permit the bearer to perform the specified activity (\texttt{\$ACTIVITY}) for any resource inside the normalized \texttt{\$PATH}.  The defined authorizations (based upon the work done in dCache for its initial Macaroon support \cite{dcache_macaroons}) are:

\begin{itemize}
    \item \texttt{UPLOAD}: Authorization to create new and upload contents, provided that existing data at the endpoint is not altered.
    \item \texttt{DOWNLOAD}: Authorization to read data.
    \item \texttt{DELETE}: Authorization to delete resources from the endpoint.
    \item \texttt{MANAGE}: Change file metadata at the storage endpoint and perform operations that may overwrite existing data.
    \item \texttt{LIST}: List the contents of a directory resource at the storage endpoint.
\end{itemize}

Beyond authorization, the experience gained in the WLCG scale tests has shown that, while HTTP-TPC can be run in either \textit{push} or \textit{pull} mode, pull mode has become preferred.  Pull mode allows the active endpoint to download with the HTTP \texttt{GET} requests; as \texttt{GET} is idempotent, the endpoint can issue numerous requests in a pipeline or partition them over a number of TCP streams to improve overall throughput.  Further, the active endpoint is the most natural entity in the system to manage a queue of transfer requests.  Given the ability to write to disk is considered a more scarce resource than reading, sites have preferred the pull mode.

\section{Conclusions}

Not only is third party copy an essential technique in the WLCG infrastructure, it is how the majority of the LHC data is transferred.  Over the past several years, the HTTP-TPC protocol has emerged as viable replacement for the venerable GridFTP protocol.  While the broad outlines of the protocol have been used by the community for years, the protocol has evolved based on operational experience and the community evolution away from using X.509 client credentials.  These activities have been led by the WLCG DOMA TPC working group which organizes several types of test-beds.
A key activity in the coming months will be to evaluate fully token-based data transfers, with authorization following the rules defined by the WLCG JWT profile~\cite{wlcg-jwt} and leveraging the integration with the WLCG IAM token issuer~\cite{wlcg-iam}.

As we go beyond tests, LHC experiments are beginning to consider the use HTTP-TPC in their production infrastructure; we expect to see significant, at-scale tests in the lead-up to Run 3.

\subsection*{Acknowledgements}

This material is based upon work supported by the National Science Foundation under Grant No. 1836650.

\bibliographystyle{plainurl}
\bibliography{main}

\begin{thebibliography}{10}

\bibitem{storm_webdav}
{The StoRM WebDAV service}.
\newblock \url{https://github.com/italiangrid/storm-webdav}.

\bibitem{wlcg-iam}
{The WLCG IAM instance}.
\newblock \url{https://wlcg.cloud.cnaf.infn.it}.

\bibitem{gfd.20}
W~Allcock.
\newblock {GridFTP: Protocol Extensions to FTP for the Grid}, April 2003.
\newblock URL: \url{https://www.ogf.org/documents/GFD.20.pdf}.

\bibitem{globus_gridfp}
{Allcock, W.}, {Bresnahan, J.}, {Kettimuthu, R.}, and {Link, M.}
\newblock {The Globus Striped GridFTP Framework and Server}.
\newblock {\em SC '05: Proceedings of the 2005 ACM/IEEE Conference on
  Supercomputing}, 2005.
\newblock \href {https://doi.org/{10.1109/SC.2005.72}}
  {\path{doi:{10.1109/SC.2005.72}}}.

\bibitem{wlcg-jwt}
{Altunay, Mine}, {Bockelman, Brian}, {Ceccanti, Andrea}, {Cornwall, Linda},
  {Crawford, Matt}, {Crooks, David}, {Dack, Thomas}, {Dykstra, David}, {Groep,
  David}, {Igoumenos, Ioannis}, {Jouvin, Michel}, {Keeble, Oliver}, {Kelsey,
  David}, {Lassnig, Mario}, {Liampotis, Nicolas}, {Litmaath, Maarten}, {McNab,
  Andrew}, {Millar, Paul}, {Sallé, Mischa}, {Short, Hannah}, {Teheran, Jeny},
  and {Wartel, Romain}.
\newblock {WLCG Common JWT Profiles (Version 1.0)}, 2019.
\newblock \href {https://doi.org/10.5281/zenodo.3460258}
  {\path{doi:10.5281/zenodo.3460258}}.

\bibitem{dcache_macaroons}
{Ashish, A}, {Millar, P.}, {Mkrtchyan, T}, {Fuhrmann, P.}, {Behrmann, G.},
  {Sahakyan, M.}, {Adeyemi1, O}, {Starek, J}, {Litvintsev, D}, and {Rossi, A}.
\newblock {dCache, towards Federated Identities \& Anonymized Delegation}.
\newblock {\em {J. Phys.: Conf. Ser.}}, 898, 2017.
\newblock \href {https://doi.org/10.1088/1742-6596/898/10/102009}
  {\path{doi:10.1088/1742-6596/898/10/102009}}.

\bibitem{fts3}
{Ayllon, A}, {Salichos, M}, {Simon, M}, and {Keeble, O}.
\newblock {FTS3: New Data Movement Service For WLCG}.
\newblock {\em J. Phys.: Conf. Ser.}, 513:032081, 2014.
\newblock \href {https://doi.org/10.1088/1742-6596/513/3/032081}
  {\path{doi:10.1088/1742-6596/513/3/032081}}.

\bibitem{macaroons}
Arnar Birgisson, Joe~Gibbs Politz, Ulfar Erlingsson, Ankur Taly, Michael
  Vrable, and Mark Lentczner.
\newblock Macaroons: Cookies with contextual caveats for decentralized
  authorization in the cloud.
\newblock In {\em NDSS}, 2014.

\bibitem{bootstrapping}
{Bockelman, Brian}, {Hanushevsky, Andrew}, {Keeble, Oliver}, {Lassnig, Mario},
  {Millar, Paul}, {Weitzel, Derek}, and {Yang, Wei}.
\newblock Bootstrapping a new {LHC} data transfer ecosystem.
\newblock {\em EPJ Web Conf.}, 214:04045, 2019.
\newblock \href {https://doi.org/10.1051/epjconf/201921404045}
  {\path{doi:10.1051/epjconf/201921404045}}.

\bibitem{rfc2068}
R.~Fielding, J.~Gettys, J.~Mogul, H.~Frystyk, and T.~Berners-Lee.
\newblock {Hypertext Transfer Protocol -- HTTP/1.1}.
\newblock Internet Requests for Comments, January 1997.
\newblock URL: \url{https://tools.ietf.org/pdf/rfc2068.pdf}.

\bibitem{globus-support}
Ian Foster.
\newblock {\em {Support for open source Globus Toolkit will end as of January
  2018}}, 2017 (accessed 28 November 2018).
\newblock URL:
  \url{https://github.com/globus/globus-toolkit/blob/4c88c9ca1423e2af806714a2eca54f6eb5d9fd4e/support-changes.md}.

\bibitem{gsi}
{Foster, Ian}, {Kesselman, Carl}, {Tsudik, Gene}, and {Tuecke, Steven}.
\newblock A security architecture for computational grids.
\newblock {\em {In Proceedings of the 5th ACM conference on Computer and
  Communications Security}}, 1998.

\bibitem{dcache_gridftp}
{Fuhrmann, P.} and {Guelzow, V.}
\newblock {dCache, Storage System for the Future}.
\newblock {\em Euro-Par 2006 Parallel Processing}, 4128, 2006.
\newblock \href {https://doi.org/10.1007/11823285_116}
  {\path{doi:10.1007/11823285_116}}.

\bibitem{httpecosystem}
{Furano, Fabrizio}, {Devresse, Adrien}, {Keeble, Oliver}, and {Hellmich,
  Martin}.
\newblock {Towards an HTTP Ecosystem for HEP Data Access}.
\newblock {\em J. Phys.: Conf. Ser.}, 2014.
\newblock URL: \url{http://iopscience.iop.org/1742-6596/513/3/032034}, \href
  {https://doi.org/10.1088/1742-6596/513/3/032034}
  {\path{doi:10.1088/1742-6596/513/3/032034}}.

\bibitem{xrootd-tpc-prot}
Andrew Hanushevsky.
\newblock {\em Third Party Copy Protocol TPC Version 2.0 Reference}, 2020
  (accessed March 7, 2020).
\newblock URL:
  \url{https://xrootd.slac.stanford.edu/doc/dev49/tpc_protocol.htm}.

\bibitem{rfc6749}
{Hardt, D.}
\newblock {The OAuth 2.0 Authorization Framework}.
\newblock Internet Requests for Comments, October 2012.
\newblock URL: \url{https://tools.ietf.org/html/rfc6749}.

\bibitem{rfc7519}
{Jones, M.}, {Bradley, J.}, and {Sakimura, N.}
\newblock {JSON Web Token (JWT)}.
\newblock Internet Requests for Comments, May 2015.
\newblock URL: \url{https://tools.ietf.org/html/rfc7519}.

\bibitem{rfc8414}
{Jones, M}, {Sakimura, N}, and {Bradley, J}.
\newblock {OAuth 2.0 Authorization Server Metadata}.
\newblock Internet Requests for Comments, June 2018.
\newblock URL: \url{https://tools.ietf.org/html/rfc8414}.

\bibitem{gfd.21}
I~Mandrichenko.
\newblock {GridFTP Protocol Improvements}, July 2003.
\newblock URL: \url{https://www.ogf.org/documents/GFD.21.pdf}.

\bibitem{rucio}
{Martin Barisits}, {Thomas Beermann}, {Frank Berghaus}, {Brian Bockelman},
  {Joaquin Bogado}, {David Cameron}, {Dimitrios Christidis}, {Diego
  Ciangottini}, {Gancho Dimitrov}, {Markus Elsing}, {Vincent Garonne},
  {Alessandro Di Girolamo}, {Luc Goossens}, {Wen Guan}, {Jaroslav Guenther},
  {Tomas Javurek}, {Dietmar Kuhn}, {Mario Lassnig}, {Fernando Lopez}, {Nicolo
  Magini}, {Angelos Molfetas}, {Armin Nairz}, {Farid Ould-Saada}, {Stefan
  Prenner}, {Cedric Serfon}, {Graeme Stewart}, {Eric Vaandering}, {Petya
  Vasileva}, {Ralph Vigne}, and {Tobias Wegner}.
\newblock Rucio: Scientific data management.
\newblock {\em Computing and Software for Big Science}, 3, 2019.
\newblock \href {https://doi.org/10.1007/s41781-019-0026-3}
  {\path{doi:10.1007/s41781-019-0026-3}}.

\bibitem{gridsite}
Andrew McNab.
\newblock {The GridSite Web/Grid security system}.
\newblock {\em {Software: Practice and Experience}}, 35:827--834, 2005.
\newblock \href {https://doi.org/10.1002/spe.690} {\path{doi:10.1002/spe.690}}.

\bibitem{smoke-tests}
Paul Millar.
\newblock {\em General utilities for working with HTTP Third-Party-Copy (TPC)},
  2020 (accessed March 7, 2020).
\newblock URL: \url{https://github.com/paulmillar/http-tpc-utils}.

\bibitem{srmv2.2}
{Sim, Alex} and {Shoshani, Arie}.
\newblock {\em The Storage Resource Manager Interface Specification Version
  2.2}, 2009 (accessed March 7, 2020).
\newblock URL: \url{https://sdm.lbl.gov/srm-wg/doc/SRM.v2.2.html}.

\bibitem{scitokens}
Alex Withers, Brian Bockelman, Derek Weitzel, Duncan Brown, Jeff Gaynor, Jim
  Basney, Todd Tannenbaum, and Zach Miller.
\newblock Scitokens: Capability-based secure access to remote scientific data.
\newblock {\em PEARC '18: Proceedings of the Practice and Experience on
  Advanced Research Computing}, 2018.
\newblock \href {https://doi.org/10.1145/3219104.3219135}
  {\path{doi:10.1145/3219104.3219135}}.

\end{thebibliography}

\end{document}